\documentclass{svjour2}                    
\usepackage{graphicx,ams}
\newcommand{\tr}{\mathop{\rm tr}}
\newcommand{\Tr}{\mathop{\rm Tr}}

\newcommand{\ci}{\mathop{\textrm{i}}}

\def\R{\mathbb R}
\begin{document}

\title{On the algebraic types of the Bel-Robinson tensor
}

\titlerunning{Types of the Bel-Robinson}        

\author{Joan Josep Ferrando \and \\ Juan Antonio S\'aez }

\authorrunning{J.J. Ferrando \and J.A. S\'aez} 

\offprints{Joan J. Ferrando}          
\institute{J.J. Ferrando \at Departament d'Astronomia i
Astrof\'{\i}sica,
\\Universitat
de Val\`encia, E-46100 Burjassot, Val\`encia, Spain.\\
\email{joan.ferrando@uv.es}   \and J.A. S\'aez \at
Departament de Matem\`atiques per a l'Economia i l'Empresa,\\
Universitat de Val\`encia, E-46071 Val\`encia, Spain.\\
\email{juan.a.saez@uv.es}}
\date{Received: date / Revised version: date}

\maketitle

\begin{abstract}
The Bel-Robinson tensor is analyzed as a linear map on the space of
the traceless symmetric tensors. This study leads to an algebraic
classification that refines the usual Petrov-Bel classification of
the Weyl tensor. The new classes correspond to degenerate type I
space-times which have already been introduced in literature from
another point of view. The Petrov-Bel types and the additional ones
are intrinsically characterized in terms of the sole Bel-Robinson
tensor, and an algorithm is proposed that enables the different
classes to be distinguished. Results are presented that solve the
problem of obtaining the Weyl tensor from the Bel-Robinson tensor in
regular cases.
\end{abstract}

\section{Introduction}
\label{intro}

With the aim of defining intrinsic states of gravitational radiation
Bel \cite{bel-1} \cite{bel-2} \cite{bel-3} introduced a rank 4
tensor which plays an analogous role for gravitational field to that
played by the energy tensor for electromagnetism. This {\it
super-energy Bel tensor} is quadratic in the Riemann tensor and, in
the vacuum case, is divergence-free. In this case the Bel tensor
coincides with the {\it super-energy Bel-Robinson tensor}, built
with the same expression by replacing the Riemann tensor with the
Weyl tensor.

Both, the Bel and the Bel-Robinson tensors have interesting
properties that, relative to every observer, allow the definition of
a non negative super-energy density and a super-energy Poynting
vector. In the last decade the interest in the super-energy tensors
has been on the increase, and a lot of works are devoted to
analyzing their properties and to studying their generalization to
any dimension and to any physical field \cite{seno} (see references
therein for an exhaustive bibliography on this subject). In a recent
work \cite{garcia}, where the dynamical laws of super-energy are
accurately analyzed, up-to-date references on the Bel and
Bel-Robinson tensors can be found.

The algebraic properties of the Bel-Robinson tensor (BR tensor) and
its close relationship with the principal null directions of the
Weyl tensor were studied by Debever early on \cite{debever-1}
\cite{debever-2}, and a spinorial approach can be found in
\cite{penri-1}. But the intrinsic algebraic characterization of a BR
tensor was not obtained until the recent work by Bergqvist and
Lankien \cite{bergqvist-lan-1}. They give the conditions on the BR
tensor playing a similar role to that played by the algebraic
Rainich \cite{rai} conditions for the electromagnetic field.

Nevertheless, some gaps remain in the algebraic comprehension of the
BR tensor. The analogy with the electromagnetic field can help us to
understand them. Let us consider an energy tensor $T_{em}$
satisfying the algebraic Rainich conditions:
\begin{equation} \label{emf-1}
T_{em}^2 = \chi^2 \, g \, , \qquad  \chi \equiv \frac12 \sqrt{\tr
T_{em}^2} \, .
\end{equation}
For a $T_{em}$ satisfying (\ref{emf-1}) we know (see for example
\cite{fsY}):

(i) Its algebraic classification: when $\chi \not= 0$, the Segr\'e
type of $T_{em}$ is $[(11)(11)]$ and it is associated with a non
null electromagnetic field $F$; when $\chi= 0$, the Segr\'e type of
$T_{em}$ is $[(31)]$ and it is associated with a null
electromagnetic field $F$.

(ii) Its canonical expression in terms of its intrinsic elements,
namely, invariant subspaces and scalars: for a non null field, $T$
takes the expression $T_{em} = - \chi \Pi$, where $\Pi$ is a 2+2
structure tensor that determines the two principal planes and the
two null principal directions of the electromagnetic field; for a
null field, $T$ takes the expression $T_{em} = l \otimes l$, where
$l$ is the fundamental vector of the electromagnetic field.

(iii) It determines the electromagnetic field $F$ up to a duality
rotation, and the explicit expression of $F$ in terms of $T_{em}$ is
also known \cite{fsY}.

As commented above, Bergqvist and Lankien \cite{bergqvist-lan-1}
have found the necessary and sufficient conditions for a rank 4
tensor $T$ to be the BR tensor associated with a Weyl-like tensor
$W$. These conditions play for the BR tensor the same role as the
Rainich conditions (\ref{emf-1}) play for the electromagnetic field.
On the other hand, the three points satated above for the
electromagnetic field could be similarly established for the BR
tensor:

(I) Algebraic classification of a BR tensor $T$: the Petrov-Bel
classification can be obtained by studying the Weyl tensor as a
linear map \cite{petrov-W}, \cite{bel-3}. The symmetries of the BR
tensor $T$ allow us to consider and analyze it as a linear map on
the nine dimensional space of the traceless symmetric tensors. What
algebraic classification follows on from this study? What
relationship exists between these BR classes and the Petrov-Bel
types of the Weyl tensor?

(II) Canonical form of the BR tensor $T$ in terms of its invariant
spaces and scalars: for every algebraic type of the BR tensor, the
eigenvalues and eigenvectors should be analyzed, as well as the
canonical expression of $T$ in terms of them. What relationship
exists between the spaces and scalar the invariants of both, the BR
tensor and the Weyl tensor?

(III) Expression of the Weyl tensor in terms of the BR tensor: it is
known that the BR tensor $T$ determines the Weyl tensor $W$ up to a
duality rotation. But the explicit expression of $W$ in terms of $T$
has not been established.

The goal of this work is to tackle the algebraic problems of the BR
tensor stated in the three points above. Here we solve the first one
and give preliminary results on the points (II) and (III) which will
be fully solved in another work in progress \cite{fs-BR-can}.

Some basic properties of the BR tensor $T$ considered as a linear
map are presented in section \ref{sec:BR-e} of this work. In section
\ref{sec:BR-p} we define the {\em main BR invariant scalars} and
study their relationship with the scalar invariants of the Weyl
tensor. This result is a contribution to point (II) above. We also
study the powers of $T$ and obtain a family of identities on the BR
tensor that generalize an already known equality. Finally, we
determine the characteristic polynomial of $T$.

In section \ref{sec-BR-Wtypes} we show that the main BR scalar
invariants allow different classes of Weyl tensors to be
distinguished. On one hand, they label the three groups of
Petrov-Bel types that can be discriminated by the Weyl eigenvalues,
namely, types III, N and O with a triple Weyl eigenvalue, types II
and D with a double eigenvalue and type I with three different
eigenvalues. On the other hand, the main BR scalars also distinguish
between five classes of algebraically general Weyl tensors. These
classes refine the Petrov-Bel classification and were introduced by
McIntosh and Arianrhod \cite{mcar}. They correspond to particular
configurations of the four null principal directions: either they
span a 3-plane or they define a frame with permutability properties.
A detailed analysis of these `degenerate' type I classes can be
found in \cite{fs-aligned-em}.

In section \ref{sec-BR-class-a} we consider the classification of
the BR tensor taking into account its eigenvalue multiplicity. We
show that seven classes appear from this point of view: one with 9
different eigenvalues, two with 6 different eigenvalues, three with
3 different eigenvalues and one with a eigenvalue of multiplicity 9.
We characterize every class in terms of the main BR invariant
scalars and show that they precisely correspond to the classes of
the Weyl tensor considered in the previous section. An arrow diagram
that helps us to visualize the different classes and their
degenerations is offered at the end of the section.

The study presented in sections \ref{sec-BR-Wtypes} and
\ref{sec-BR-class-a} shows that the main BR scalars do not
distinguish between Petrov-Bel types III, N and O, and between types
II and D. Nevertheless we prove in section \ref{sec-BR-class-b} that
all these Petrov-Bel types match with different classes of the BR
tensor and that they can be discriminated by using their minimal
polynomial. At the end of this last section we present an algorithm
that summarizes the classification of the BR tensor and the
intrinsic characterization of every class.

In studying the minimal polynomial in last section we use a
expression that relates the $n$-powers of the BR and Weyl tensors.
This expression is proved in section \ref{sec-WfromBR} and it is
valid except for some degenerate cases. For the case $n=3$ this
expression allows us to obtain, in the same section, the Weyl tensor
in terms of the BR tensor when none of the main BR scalar invariants
vanish. This result partially solves point (III) above. Elsewhere
\cite{fs-BR-can} we obtain the Weyl tensor in terms of the BR tensor
in the cases that are not considered here.

With the intention of making the mathematical expressions simpler we
use, when possible, a global notation that is accurately explained
in four appendixes at the end of this work. This also allow us to
simplify the calculations by using some properties that are also
stated in these appendixes.

It is worth outlining that, in spite of the fact that the BR tensor
loses part of the information of the Weyl tensor (a duality
rotation), its study as a linear map leads to ten different classes,
a richer classification than the six Petrov-Bel types of the Weyl
tensor as a linear map. On the other hand, note that from our study
follows a characterization of every Petrov-Bel type in terms of the
BR tensor, a result that only was known for types N and O
\cite{bose} \cite{bergqvist}.

\section{The Bel-Robinson tensor: towards its algebraic classification}
\label{sec:BR-e}

We shall note $g$ the space-time metric with signature $\{ -, +,+,+
\}$ and we shall write $W$ to indicate the Weyl tensor. The
conventions of signs are those of the book by Stephani {\em et al.}
\cite{kra}. The Bel-Robinson tensor (BR tensor) $T$  is given in
terms of the Weyl tensor as:
\begin{equation} \label{BR-1}
{T_{\alpha  \mu \beta \nu}} = \frac14 \left({{{
W_{\alpha}}^{\rho}}_{\beta}}^{ \sigma} W_{\mu \rho \nu \sigma } +
{{{* W_{\alpha}}^{\rho}}_{\beta}}^{ \sigma} * W_{\mu \rho \nu \sigma
}\right) \, ,
\end{equation}
where $*$ denotes the Hodge dual operator (see Appendix
\ref{ap-SD}.9). Note that expression (\ref{BR-1}) coincides with
that given in \cite{penri-1} \cite{bergqvist} and it differs in a
factor from the original one by Bel \cite{bel-3}.

\subsection{The Weyl tensor as an endomorphism}
\label{WT-e}

A self--dual 2--form is a complex 2--form ${\cal F}$ such that
$*{\cal F}= \textrm{i}{\cal F}$. We can associate biunivocally with
every real 2--form $F$ the self-dual 2--form ${\cal
F}=\frac{1}{\sqrt{2}}(F-\textrm{i}*F)$. In short, here we refer to a
self--dual 2--form as a {\it SD bivector}. The endowed metric on the
3-dimensional complex space of the SD bivectors is ${\cal
G}=\frac{1}{2}(G-\textrm{i} \; \eta)$, $\eta$ being the metric
volume element of the space-time, and $G$ being the metric on the
space of 2--forms, $G=\frac{1}{2} g \wedge g$. Here $\wedge$ denotes
the double-forms exterior product (see Appendix
\ref{ap-2tensors}.5).

The algebraic classification of the Weyl tensor $W$ can be obtained
\cite{petrov-W}, \cite{bel-3} by studying the traceless linear map
defined by the self--dual (SD) Weyl tensor ${\cal W}=\frac{1}{2}
(W-\textrm{i}*W)$ on the SD bivectors space. This {\em
SD-endomorphism} (see notation in Appendix \ref{ap-SD}) has
associated the complex scalar invariants:
\begin{equation} \label{weylinva}
a = \Tr {\cal W}^2 , \quad b = \Tr {\cal W}^3 \, .
\end{equation}
In terms of them, the characteristic equation reads
\begin{equation} \label{eccar-weyl-1}
x^{3}-\frac{1}{2} ax -\frac{1}{3} b =0 \, .    \label{ce}
\end{equation}
Then, the Petrov-Bel classification follows taking into account both
the eigenvalue multiplicity and the degree of the minimal
polynomial. The algebraically regular case (type I) occurs when the
characteristic equation (\ref{ce}) admits three different roots. If
there is a double root $\rho= -{b \over a}$ and a simple one
$-2\rho$, the minimal polynomial distinguishes between types D and
II. Finally, if all the roots are equal, and so zero, the Weyl
tensor is type O, N or III, depending on the minimal polynomial.

\subsection{{The Bel-Robinson tensor as an endomorphism}}
\label{BRT-e}

The expression (\ref{BR-1}) of the BR tensor may be written in terms
of the SD Weyl tensor as:
\begin{equation}\label{BR-2}
{T_{\alpha  \mu \beta \nu}} = {{{{\cal
W}_{\alpha}}^{\rho}}_{\beta}}^{ \sigma} \overline{\cal W}_{\mu \rho
\nu \sigma }  \, ,
\end{equation}
where $ \bar{\ } $ stands for complex conjugate. Taking into account
the $\diamond$-product defined in Appendix \ref{ap-SD-TLS}, we have
that (\ref{BR-2}) becomes
\begin{equation} \label{BR-3}
T = {\cal W} \diamond \overline{\cal W}  \, .
\end{equation}
It is easy to show that, for two arbitrary SD-endomorphisms ${\cal
X}$ and ${\cal Y}$, the tensor $Q = {\cal X} \diamond \overline{\cal
Y}$ satisfies
\begin{equation} \label{TLS-1}
Q_{\alpha  \mu \beta \nu} = Q_{\mu \alpha \beta \nu} = Q_{\beta \nu
\alpha \mu} \, , \qquad  Q_{\ \alpha  \beta \nu}^{\alpha} = 0 \, .
\end{equation}
As a consequence of these properties, if $S$ is a trace-less
symmetric tensor (TLS tensor), then $Q(S)_{\alpha \beta} =
{Q_{\alpha \beta}}^{ \mu \nu} \ S_{\mu \nu}$ is trace-less and
symmetric. Thus, we can see $Q$ as a (complex) symmetric linear map
on the space of the TLS tensors, that is to say, $Q$ is a {\em
TLS-endomorphism} (see notation in Appendix \ref{ap-TLS}).

Besides, the BR tensor as given by (\ref{BR-3}) has two important
additional properties. Firstly, it is real and, secondly, as a
consequence of the traceless property of the SD Weyl tensor, it is a
fully symmetric traceless tensor (see Appendix \ref{ap-SD-TLS}.7).
Then we have:
\begin{lemma}
The Bel-Robinson tensor $T$ is a real traceless TLS-endomorphism,
that is, a traceless symmetric endomorphism on the 9-dimensional
space of the traceless symmetric tensors.
\end{lemma}

As explained in subsection above, the Petrov-Bel classification
follows on by studying the Weyl tensor as a SD-endomorphism.
Similarly, we want to obtain the algebraic classification of the BR
tensor by analyzing it as a TLS-endomorphism. We start our study by
writing the BR tensor in terms of another TLS-endomorphism. If we
define
\begin{equation} \label{Omega-1}
\Omega = {\cal W} \diamond \overline{\cal G}  \, ,
\end{equation}
and we take into account (\ref{BR-3}) and that $T$ is real, we
obtain:
\begin{equation}\label{BR-4}
T= \Omega \bullet \overline{\Omega}  = \overline{\Omega} \bullet
\Omega \, ,
\end{equation}
where $\bullet$ denotes the composition of TLS-endomorphisms (see
Appendix \ref{ap-TLS}).

The tensor $\Omega$ is not but the restriction of the Weyl tensor
acting (with the indices 2 and 4) on the TLS tensor space. Indeed,
for a trace-less symmetric tensor $S$, we have
\begin{equation} \label{Omega-2}
{\Omega^{\alpha \beta}}_{\mu \nu} \ S^{\mu \nu} = {\cal W}^{\,
\alpha  \ \, \beta}_{\ \, \mu \ \ \nu} \ S^{\mu \nu}  \, .
\end{equation}
On the other hand, it is easy to show that the tensorial components
of $\Omega$ are
\begin{equation}\label{Omega-3}
\Omega_{\alpha \beta \mu \nu}= - {\cal W}_{\alpha (\mu \nu) \beta}
\, ,
\end{equation}
where $(\ \ )$ denotes symmetrization.

The tensor $\Omega$ inherits some properties of the SD Weyl tensor.
Evidently, it is a TSL-endomorphism (it satisfies (\ref{TLS-1}))
and, moreover, it is traceless and satisfies a SD-like property:
\begin{equation} \label{Omega-4}
\Omega^{\alpha}_{\  \mu \alpha \nu} = 0 \, , \qquad \Omega_{\alpha
\gamma \beta \delta} \, {\overline{\cal G}^{\gamma \delta}}_{ \mu
\nu} = 0 \, .
\end{equation}

Conversely, if $\Omega$ is a (complex) TLS-endomorphism that
satisfies (\ref{Omega-4}) we can reverse the expression
(\ref{Omega-1}) and obtain the traceless SD-endomorphism (see
property of Appendix \ref{ap-SD-TLS}.5):
\begin{equation} \label{Omega-6}
{\cal W}_{\alpha \beta \mu \nu} = \frac23 \, \Omega_{\alpha \gamma
\beta \delta} \, {{\cal G}^{\gamma \delta}}_{ \mu \nu} = \frac43
\Omega_{\alpha [\mu \nu] \beta} \, ,
\end{equation}
where $[\ \ ]$ denotes antisymmetrization. Thus we have the
following.
\begin{proposition} There is a one-to-one correspondence between
traceless SD-endomorphisms ${\cal W}$  and the (complex)
TLS-endomorphism $\Omega$ which satisfy {\rm (\ref{Omega-4})}. This
bijection is given by {\rm (\ref{Omega-1})} and {\rm
(\ref{Omega-6})}.
\end{proposition}

\subsection{Can the Weyl tensor be obtained from the Bel-Robinson tensor?}
\label{subsec-W-BR}

The BR tensor can be obtained from the Weyl tensor by means of the
quadratic expression (\ref{BR-1}), but it is known that this
expression is invariant under duality rotation of the Weyl tensor.
For the SD Weyl tensor a duality rotation takes the form $e^{\ci
\theta} {\cal W}$. Then, the above quoted invariance follows
trivially from the expression (\ref{BR-3}) because,
$${\cal W} \diamond \overline{\cal W} =[ e^{\ci
\theta} {\cal W}] \diamond [e^{- \ci \theta} \overline{\cal W}]   \,
.$$

Thus, the accurate question we should pose is: can the Weyl tensor
be determined, up to a duality rotation, from the BR tensor? or, to
be more precise, is there an explicit algorithm to obtain it? These
questions will be analyzed and solved elsewhere \cite{fs-BR-can} for
an arbitrary BR tensor. In section \ref{sec-WfromBR} we study this
question for regular cases. In this study the TLS endomorphism
$\Omega$ plays an important role.

\section{Powers, scalar invariants and characteristic polynomial
of the Bel-Robinson tensor} \label{sec:BR-p}

In order to tackle the algebraic classification of the BR tensor $T$
as a TLS-endomorphism we need to know its characteristic polynomial
and its scalar invariants. In this section we obtain them from the
powers of $T$.

The powers of the BR tensor can be obtained as $T^n = T^{n-1}
\bullet T$. From (\ref{BR-4}) and as $T$ is real, the $\Omega$
defined in (\ref{Omega-1}) satisfies $\Omega \bullet
\overline{\Omega} = \overline{\Omega} \bullet \Omega $. Then, from
(\ref{BR-4}) we obtain
\begin{lemma} For $n \geq 1$ it holds $T^n = \Omega^n \bullet  {\overline{\Omega}^n}$,
where $\Omega$ is the endomorphism defined as {\rm (\ref{Omega-1})}
from the Weyl tensor ${\cal W}$.
\end{lemma}

On the hand we can obtain the TLS-endomorphism $\Omega^{n} = \Omega
\bullet \Omega^{n-1}$ in terms of the SD-endomorphism ${\cal W}^n =
{\cal W} \circ {\cal W}^{n-1}$. Indeed, from the definition
(\ref{Omega-1}) and putting, in expression (\ref{ap-D4}) of Appendix
\ref{ap-SD-TLS}.8, ${\cal V} = {\cal W}$, ${\cal X} = {\cal
W}^{n-1}$ and ${\cal Z} = {\cal Y} = {\cal G}$ , we obtain by
induction:
\begin{lemma} \label{lemma-Omega-n}
The TLS-endomorphism $\Omega$ given by {\rm (\ref{Omega-1})}
satisfies
\begin{equation} \label{omegaene}
\Omega^{n} =  \  {\cal W}^n \diamond \overline{\cal G}, \qquad n
\geq 1  \, .
\end{equation}
\end{lemma}

The expression for the powers of the BR tensor and the traces of
these powers in terms of the SD Weyl tensor can also be obtained by
induction by taking, in expression (\ref{ap-D4}) of Appendix
\ref{ap-SD-TLS}.8, ${\cal V} = {\cal Z} = {\cal W}$, ${\cal X} =
{\cal Y} = {\cal W}^{n-1}$, and taking into account the property of
Appendix \ref{ap-SD-TLS}.2. Thus, we have:
\begin{proposition} \label{prop-Tn-trTn}
The powers of the BR tensor and their traces can be computed in
terms of the SD Weyl tensor as
\begin{equation} \label{Tn-trTn}
T^n =  {\cal W}^n \diamond \overline{\cal W}^n \, , \qquad \tr T^n =
| \tr {\cal W} ^n |^2  \, .
\end{equation}
\end{proposition}
From this proposition, and by applying the property of Appendix
\ref{ap-SD-TLS}.2, we obtain the following result easily:
\begin{corollary}
The powers $T^n$ of the BR tensor satisfy:
\begin{equation} \label{involutiu}
(T^n)^{ \lambda}_{\ \alpha \lambda \beta } = \frac14 \Tr T^n \,
g_{\alpha \beta}  \, .
\end{equation}
\end{corollary}
For $n=2$ this expression states that $T_{\alpha \lambda \mu \nu}
{T_{\beta}}^{\lambda \mu \nu} = \frac14 T_{\rho \lambda \mu \nu}
T^{\rho \lambda \mu \nu} g_{\alpha \beta}$, an identity already
known \cite{debever-1} \cite{penri-1}. Thus (\ref{involutiu})
generalizes, for an arbitrary $n$, this identity.

As a consequence of proposition \ref{prop-Tn-trTn}, the traces of
the powers of $T$ are non negative and we can compute them in terms
of the traces of the SD Weyl tensor. But we also know that ${\cal
W}$ satisfies the characteristic equation (\ref{eccar-weyl-1}), that
is:
\begin{equation} \label{eccar-weyl-2}
{\cal W}^3 =  \frac{1}{2} a {\cal W} + \frac{1}{3} b {\cal G}  \, ,
\end{equation}
where $a$ and $b$ are the {\em main Weyl scalar invariants} given in
(\ref{weylinva}).

From (\ref{eccar-weyl-2}), the traces of the powers of ${\cal W}^n$
for $n>3$ can be computed in terms of $a$ and $b$. After that, they
can be used to compute the traces of $T^n$ by applying proposition
\ref{prop-Tn-trTn}. More precisely, if we denote
\begin{equation} \label{belab}
\alpha = \frac{1}{2} | a | ,   \qquad  \beta = \frac{1}{3} | b |,
\qquad \mu = \frac{1}{2^3 3^2} \Big( a^3 \bar{b}^2 + \bar{a}^3 b^2
\Big)  \, ,
\end{equation}
we have that the traces of $T^n$ are given by
\begin{equation} \label{traces}
\begin{array}{ll}
\Tr T= 0 \,   & \Tr T^2 = 4 \alpha^2 \,  \\   \Tr T^3 = 9 \beta^2 \,
 & \Tr T^4 = 4   \alpha^4 \,  \\  \Tr T^5 = 25 \alpha^2 \beta^2 \,&
\Tr T^6 = 4  \alpha^6 + 9  \beta^4 + 6 \mu  \, \\
\Tr T^7 = 49  \alpha^4  \beta^2 \,  \qquad & \Tr T^8 = 4  \alpha^8 +
64   \alpha^2   \beta^4  + 12   \alpha^2 \mu  \, \\
 \Tr T^9 = 9 \beta^2 [ \beta^4 + 9 \alpha^6 + 3 \mu ]  \, .\quad &
\end{array}
\end{equation}

Note that the Weyl tensor defines four real scalar invariants that
can be grouped in the complex ones $a$, $b$ defined in
(\ref{weylinva}) from the SD Weyl tensor. We have remarked in
subsection \ref{subsec-W-BR} that a duality rotation is lost when
constructing the BR tensor. So just three scalars survive in $T$: we
can see in (\ref{traces}) that two of them, the modulus of $a$ and
$b$ determine the traces of $T^2$ and $T^3$. The third one, $a^3
\bar{b}^2 + \bar{a}^3 b^2$, does not appear until $\Tr T^6$ is
computed. This fact and expressions (\ref{traces}) justify defining
the following {\em main BR scalar invariants}:
\begin{equation} \label{invar-BR}
\alpha = \frac{1}{2} \sqrt{\Tr T^2} , \qquad \beta = \frac{1}{3}
\sqrt{\Tr T^3} \, , \qquad \mu = \frac{1}{6} \Big( \tr T^6 - 4
\alpha^6 - 9 \beta^4 \Big)  \, .
\end{equation}

Then, the traces of the powers of the BR tensor depend on the BR
main scalar invariants as (\ref{traces}). Moreover, these
expressions can be used to obtain the characteristic equation of the
BR tensor and we achieve the following.
\begin{proposition}  \label{propcar}
The BR tensor satisfies the characteristic equation
\begin{equation} \label{charac-polyno}
x^9 - 2 \alpha^2 x^7 - 3   \beta^2 x^6 +  \alpha^4 x^5 + \alpha^2
\beta^2 x^4 + ( 3 \beta^4 - \mu) x^3 +   \alpha^2 \beta^4 x -
\beta^6 = 0\end{equation}
where $\alpha$, $\beta$ and $\mu$ are the
main BR scalar invariants {\rm (\ref{invar-BR})}. These invariants
determine the traces $\Tr T^n$ as {\rm (\ref{traces})}.
\end{proposition}

\section{Labeling some Weyl types with the main BR scalar invariants}
\label{sec-BR-Wtypes}

In the next section we study the algebraic types of the BR tensor as
TLS-endomorphism. In order to understand the correspondence between
these new BR types and the already known Weyl types, we show in this
section that some of these Weyl classes can be characterized in
terms of the main BR scalar invariants.

When analyzing the Weyl tensor as a SD-endomorphism, the main Weyl
invariant scalars distinguish the multiplicity of the Weyl
eigenvalues: the algebraically general type I has 3 simple
eigenvalues, types II and D have a simple and a double one, and
types III, N and O have a triple vanishing eigenvalue. Then, from
the characteristic equation (\ref{eccar-weyl-1}) we obtain
\cite{kra} \cite{fms}:
\begin{lemma} \label{lemma-petrov-bel}
Let $a = \Tr {\cal W}^2$, $b = \Tr {\cal W}^3$ be the main Weyl
scalar invariants. Then, the Weyl tensor is

i. Petrov-Bel type I iff $\ 6 b^2 - a^3 \neq 0$.

ii. Petrov-Bel types D or II iff $\ 6 b^2 = a^3 \neq 0$.

iii. Petrov-Bel types O, N or III iff $\ a=0=b$.

\end{lemma}
Note that in order to distinguish between types D and II as well as
between types N or III, the minimal polynomial is necessary
\cite{fms}. But, at the moment, we are only interested in scalar
conditions.

On the other hand, some classes of algebraically general space-times
have been considered in literature. A refinement of the Petrov-Bel
classification is based on the Weyl scalar invariant \cite{mcar}:
\begin{equation} \label{eme}
M= \frac{a^3}{b^2} - 6 \, .
\end{equation}
The invariant $M$ is not defined if $b=0$ as occurs in types III and
N and in algebraically general space-times with a vanishing
eigenvalue. But we can extend its validity to these cases if we put
$M= \infty$ whenever $b=0 \neq a$, and $M=0$ if $a=0=b$.

This way, it is known \cite{mcar} \cite{fsI} that the cases where
$M$ is real positive or infinity, which are called IM$^+$ or
IM$^\infty$ respectively, correspond to the case of the Weyl
eigenvalues having a real ratio, $\frac{\rho_i}{\rho_j} \in \R $
(the case $M= \infty$ means that the Weyl tensor has a vanishing
eigenvalue). An equivalent condition in terms of Debever null
directions has also been obtained \cite{mcar} \cite{fsI}: M is real
positive or infinity iff the four Debever null directions span a
3-plane.

On the other hand, the case where $M$ is real negative  IM$^-$ has
also been considered. It corresponds to the property of two of the
SD eigenvalues having the same modulus ($| \rho_i | = |\rho_j |$ for
some $i \neq j$). Moreover, the case $M=-6$ (or $a=0$) corresponds
to all the eigenvalues having the same modulus. These conditions
have also been interpreted in terms of permutability properties of
the null Debever directions \cite{fs-aligned-em}.

A detailed analysis of these 'degenerate' algebraically general
classes can be found in \cite{fs-aligned-em}, where they are also
interpreted as the space-times where the electric and magnetic parts
of the Weyl tensor are aligned for a (non necessary time-like)
direction. These classes also contain the purely electric and purely
magnetic space-times which have been accurately studied in
\cite{fsWem}.

The scalar $M$ is homogeneous in the Weyl tensor and thus invariant
by duality rotation. Consequently, it should depend on the main BR
scalars. Indeed, if we take into account the relations (\ref{belab})
between the main Weyl scalars $a$, $b$ and the main BR scalars
$\alpha$, $\beta$ and $\mu$, we obtain:
$$
81 \, \beta^8 \, (Im[M])^2 = 16 \, (4 \alpha^6 \beta^4 - \mu^2) \, ,
\qquad 9 \, \beta^4 \, Re[M]  = 4 \, \mu -  54 \, \beta^4 \, .
$$
The first expression shows that $M$ is real if, and only if, $\mu^2-
4 \alpha^6 \beta^4 =0$. After that, if $M$ is real, the second
expression gives its sign in terms of the main BR scalars. More
precisely, we have:
\begin{lemma} \label{lemma-BR-scalars}
The invariant scalar $M$ is real, if and only if $\mu^2- 4 \alpha^6
\beta^4 =0$. If this condition holds, the Weyl tensor is:

i. Type $IM^+$ ($M$ real positive) iff $\ 2 \mu - 3^3 \beta^4 >0$.

ii. Type $IM^\infty$  ($M = \infty$) iff $\ \beta =0 \neq \alpha$.

iii. Type $IM^-$  ($M$ real negative) iff $\ 2 \mu - 3^3 \beta^4 <
0$

iv. Type $IM^{[-6]}$  ($M =-6$) iff $\ \alpha=0 \neq \beta$.

v. Algebraically special ($M=0$) iff $\ 2 \mu - 3^3 \beta^4 = 0$
\end{lemma}

From now on we denote $I_r$ the class of non `degenerate' type I
Weyl tensors, that is to say, those with non real invariant $M$.
Then, from lemmas \ref{lemma-petrov-bel} and \ref{lemma-BR-scalars}
and relations (\ref{belab}) we obtain the following characterization
of some classes of the Weyl tensor in terms of the main BR scalars:
\begin{proposition} \label{prop-BR-scalars}
If $\alpha$, $\beta$ and $\mu$ are the main BR invariant scalars,
then the Weyl tensor is:

i. Type O, N or III iff $\ \alpha = \beta =0$.

ii. Type D or II iff  $\ \mu^2- 4 \alpha^6 \beta^4 =0$, $\ 2^2
\alpha^3 = 3^3 \beta^2 \neq 0 $.

iii. Type $IM^+$ iff   $\ \mu^2- 4 \alpha^6 \beta^4 =0$, $\ 2 \mu -
3^3 \beta^4 >0$.

iv. Type $IM^\infty$ iff $\ \alpha \neq 0 = \beta$.

v. Type $IM^-$ iff $\ \mu^2- 4 \alpha^6 \beta^4 =0$, $\ 2 \mu - 3^3
\beta^4 <0$.

vi. Type $IM^{[-6]}$ iff $\ \beta \neq 0 = \alpha$.

vii. Type $I_r$ iff $\ \mu^2- 4 \alpha^6 \beta^4 \not= 0$.

\end{proposition}

\section{Classifying the Bel-Robinson tensor}
\label{sec-BR-class-a}

In this section we study the BR tensor as a TLS-endomorphism taking
into account the eigenvalue multiplicity.

As a consequence of (\ref{Omega-1}), $\Omega$ and ${\cal W}$ satisfy
the same minimal polynomial, that is, both have the same eigenvalues
$\rho_i$, $i=1,2,3$. On the other hand, from expression
(\ref{BR-4}), $\Omega$ and $\overline{\Omega}$ are symmetric
endomorphisms that commute and, consequently, they have the same
eigenvectors. These eigenvectors and their associated eigenvalues
should be either real or pairs of complex conjugates and then $T$
will have the same eigenvectors with associated eigenvalues $\rho_i
\bar{\rho}_j$. Thus, we obtain the BR eigenvalues without solving
the characteristic equation (\ref{charac-polyno}).
\begin{proposition}
The BR tensor has 3 real eigenvalues $t_i$, and 3 pairs of complex
conjugate eigenvalues $\tau_i$, $\bar{\tau}_i$. They depend on the
Weyl eigenvalues $\rho_i$ as
\begin{equation} \label{BR-eigenvalues-1}
t_i =| \rho_i |^2  \, ; \qquad \tau_i = \rho_j \bar{\rho}_k \, ,
\quad  (i,j,k) \ \textrm{a pair permutation of } \, (1,2,3) \, .
\end{equation}
\end{proposition}

The BR tensor has three main scalar invariants. In accordance with
this fact, the BR complex eigenvalues should depend on the three
real ones. Indeed, a straightforward calculation shows
\begin{lemma}
In terms of the real BR eigenvalues $t_i$ the complex ones $\tau_k$
take the expression:
\begin{equation} \label{BR-eigenvalues-2}
\tau_k = p_k + \ci \,   q \, , \quad 2 p_k \equiv t_k - t_i - t_j \,
, \ (i,j,k \neq) \, ,  \quad q^2 \equiv p_1 p_2 + p_2 p_3 + p_3 p_1
\, .
\end{equation}
\end{lemma}

Generically, the BR eigenvalues are nine different scalars. Now we
analyze the admitted degenerations. If $t_i = t_j \not= t_k$, then
(\ref{BR-eigenvalues-2}) implies $p_i = p_j$, that is, $\tau_i =
\tau_j$ and $\bar{\tau}_i = \bar{\tau}_j$ and, consequently, $T$ has
3 double eigenvalues. Moreover, expression (\ref{BR-eigenvalues-1})
implies that the Weyl tensor has two eigenvalues with the same
modulus, $|\rho_i | = | \rho_j |$, a case corresponding to a Weyl
tensor of type $IM^-$. If we avoid a higher degeneration by
considering $t_k \not = t_i = t_j$, then $M \not= -6$.

If one of the complex eigenvalues $\tau_i$ is real, we have $\tau_i
= \bar{\tau}_i$. Then (\ref{BR-eigenvalues-2}) implies $q=0$ and
consequently, $\tau_k = \bar{\tau}_k$ for every $k=1,2,3$. Thus $T$
has, again, 3 double eigenvalues. Now, from
(\ref{BR-eigenvalues-1}), the ratio of the Weyl eigenvalues
$\frac{\rho_i}{\rho_j}$ is real for every pair $(i,j)$ because
$\rho_i \bar{\rho}_j$ is real. If we avoid higher degenerations we
must remove the case of a vanishing Weyl eigenvalue ($M
\not=\infty$) and then the Weyl tensor is type $IM^+$.

Thus, the case of nine different eigenvalues corresponds to a
regular type $I_r$ Weyl tensor. From all these considerations and
taking into account proposition \ref{prop-BR-scalars} we have the
followings.

\begin{proposition} \label{prop-BR-eignevalues-a-0}
A BR tensor $T$ has nine different eigenvalues (3 real and 3 pairs
of complex conjugate) iff $\ \mu^2- 4 \alpha^6 \beta^4 \neq 0$, that
is to say, the Weyl tensor is type $I_r$.
\end{proposition}

\begin{proposition} \label{prop-BR-eignevalues-a}
If $\ \mu^2- 4 \alpha^6 \beta^4 = 0$, the BR tensor $T$ has, at the
most, six different eigenvalues. Then we have two cases:

(i) $T$ has a double and a simple real eigenvalues, and two double
and two simple complex conjugate eigenvalues iff $\ 2 \mu - 3^3
\beta^4 <0$ and $\ \alpha \not=0$, that is to say, the Weyl tensor
is type $IM^-$ with $M \not= -6$.

(ii) $T$ has three simple and three double real eigenvalue iff $\ 2
\mu - 3^3 \beta^4 >0$ and $\ \beta \not=0$, that is to say, the Weyl
tensor is type $IM^+$.
\end{proposition}

Starting from the two cases above with six different eigenvalues we
can consider three kinds of further degeneration. The first one
follows on by imposing both conditions, $t_i = t_j \not= t_k$ and
$\tau_i = \bar{\tau}_i$, which is a degeneration of both, the $IM^-$
($M \not= -6$) and $IM^+$ types. Then, from (\ref{BR-eigenvalues-2})
we have $\tau_i = \bar{\tau}_i = \tau_j = \bar{\tau}_j$ and $\tau_k
= t_i$ and, consequently, $T$ has four real eigenvalues with
multiplicities 4, 1, and 4. Moreover, (\ref{BR-eigenvalues-1})
implies that the Weyl tensor has two equal eigenvalues, $\rho_i =
\rho_j$ and, consequently, it is Petrov-Bel type D or II.

Secondly, we can consider a degeneration of the $IM^-$ type by
taking the three real eigenvalues as equal, $t_1 = t_2= t_3$. Then
(\ref{BR-eigenvalues-2}) implies $\tau_1 = \tau_2= \tau_3$ and $T$
has three different eigenvalues with multiplicity three, a real one
and a pair of complex conjugates. On the other hand, if we avoid a
further degeneration, (\ref{BR-eigenvalues-1}) implies that $M=-6$.

Thirdly, we can consider a degeneration of the $IM^+$ type by
imposing $\tau_i = t_j$. Then (\ref{BR-eigenvalues-2}) implies $t_j
= \tau_i = \tau_k = 0$ $(i,j,k \neq)$,  and $T$ has three real
eigenvalues with multiplicities 5, 2 and 2. From
(\ref{BR-eigenvalues-1}) we have $\rho_j=0$ and then $M= \infty$.

Note that, as a consequence of (\ref{BR-eigenvalues-2}), imposing
$\tau_i = t_i$ to the $IM^+$ case leads to vanishing eigenvalues, a
more degenerate case that we will consider below. All these
considerations and proposition \ref{prop-BR-scalars} lead to the
following.

\begin{proposition}  \label{prop-BR-eignevalues-b}
(i) A BR tensor $T$ has three real eigenvalues with multiplicities
4, 1, and 4 iff $\ \mu^2- 4 \alpha^6 \beta^4 = 0$ and $\ 2^2
\alpha^3 = 3^3 \beta^2 \neq 0 $, that is to say, the Weyl tensor is
type $II$ or $D$.

(ii) A BR tensor $T$ has three triple eigenvalues (a real and a pair
of complex conjugates) iff $\ \beta \neq 0 = \alpha$, that is to
say, the Weyl tensor is type $IM^{[-6]}$.

(iii) A BR tensor $T$ has three real eigenvalues (with
multiplicities 2, 5 and 2) iff  $\ \beta = 0 \neq \alpha$, that is
to say, the Weyl tensor is type $IM^{\infty}$.
\end{proposition}

Finally we can consider a further degeneration from the three cases
in proposition above. By using (\ref{BR-eigenvalues-2}) it is easy
to show that in any case we reach the highest degeneration: there is
a unique vanishing eigenvalue. Then, (\ref{BR-eigenvalues-1})
implies that the Weyl eigenvalues also vanish and, consequently, the
Weyl tensor is Petrov-Bel type III, N or O. Thus, from proposition
\ref{prop-BR-scalars} we have:

\begin{proposition}  \label{prop-BR-eignevalues-c}
A BR tensor $T$ has a sole vanishing eigenvalue with multiplicity 9
iff $\ \alpha = \beta =0$, that is to say, the Weyl tensor is type
$III$, $N$ or $O$.
\end{proposition}

It is worth remarking that this proposition implies that the
Petrov-Bel types $III$, $N$ and $O$ can not be distinguished by
taking into account the eigenvalue multiplicity of the BR tensor $T$
or, equivalently, by analyzing its main scalar invariants. We find a
similar situation for the Petrov-Bel types $II$ and $D$ as a
consequence of the first point in proposition
\ref{prop-BR-eignevalues-b}. Nevertheless, we will see in the last
section that each of these Petrov-Bel types corresponds to a
different algebraic BR type and that they can be distinguished by
using the minimal polynomial of $T$ as a TLS-endomorphism.

Propositions \ref{prop-BR-eignevalues-a},
\ref{prop-BR-eignevalues-b} and \ref{prop-BR-eignevalues-c} classify
and characterize the BR tensor taking into account its eigenvalue
multiplicity. In order to better visualize the eigenvalue
degeneration that relates the different classes, we present the
following arrow diagram. Note that the four files in the diagram
correspond to 9, 6, 3 and 1 different eigenvalues.

\vspace*{0.3cm}

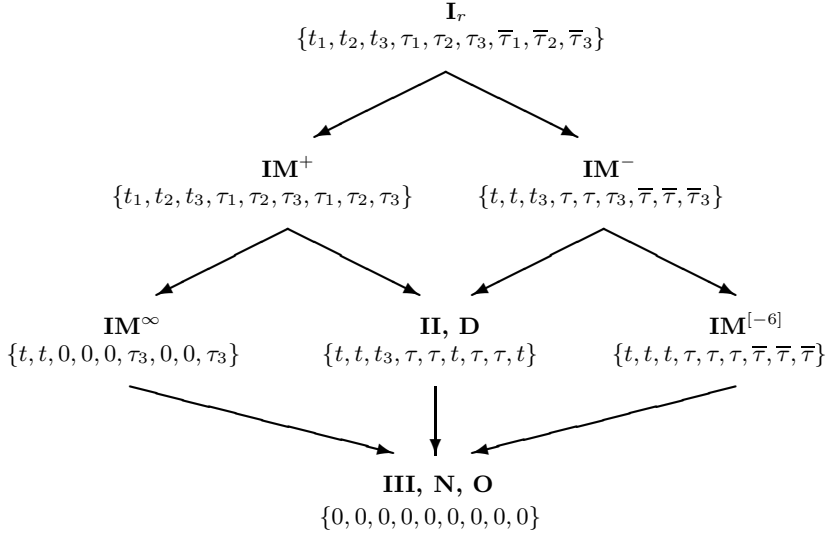
 \setlength{\unitlength}{0.7cm} {\footnotesize \noindent
\begin{picture}(0,11)
\thicklines

\put(9,10){\bf{I$_r $}}

 \put(6,9.5){ $\left\{ t_1, t_2 , t_3 , \tau_1 , \tau_2 ,
\tau_3 , \overline{\tau}_1 , \overline{\tau}_2 , \overline{\tau}_3
\right\}$}

\put(9,9){\vector(-2,-1){2.5}} \put(9,9){\vector(2,-1){2.5}}

\put(5.5,7){\bf{IM$^+ $}}

\put(2.5,6.5){ $\left\{ t_1, t_2 , t_3 , \tau_1 , \tau_2 , \tau_3 ,
{\tau}_1 , {\tau}_2 , {\tau}_3 \right\} $}

\put(11.6,7){\bf{IM$^- $ }}

\put(9.6,6.5){$\left\{ t, t , t_3 , \tau , \tau , \tau_3 ,
\overline{\tau} , \overline{\tau} , \overline{\tau}_3 \right\}$}

\put(6,6){\vector(-2,-1){2.5}} \put(6,6){\vector(2,-1){2.5}}
\put(12,6){\vector(-2,-1){2.5}} \put(12,6){\vector(2,-1){2.5}}

\put(2.5,4){\bf{IM$^\infty $ }} \put(14,4){\bf{IM$^{[-6]} $ }}
\put(8.5,4){\bf{II, D}}

\put(0.5,3.5){ $\left\{ t, t ,0, 0, 0 , \tau_3 , 0, 0, \tau_3
\right\} $}

\put(6.5,3.5){ $\left\{t, t, t_3  , \tau  , \tau , t, \tau  , \tau ,
t \right\} $}

\put(12,3.5){ $ \left\{t, t, t  , \tau  , \tau , \tau,
\overline{\tau} ,\overline{\tau}  ,\overline{\tau} \right\} $}

\put(8.8,3){\vector(0,-1){1.3}} \put(14.5,3){\vector(-4,-1){5}}
\put(3,3){\vector(4,-1){5}}

\put(7.8,1){\bf{III, N, O }}

\put(6.6,0.4){$\{ 0,0,0,0,0,0,0, 0, 0 \}$}

\end{picture}}


\section{Obtaining the Weyl tensor from the BR tensor: regular case}
\label{sec-WfromBR}

As commented in subsection \ref{subsec-W-BR} the general problem of
obtaining the Weyl tensor, up to a duality rotation, in terms of the
BR tensor will be solved elsewhere \cite{fs-BR-can}. The specific
algorithm strongly depends on the algebraic type of the Weyl and BR
tensors. Here we present an explicit expression which is valid for
the Petrov-Bel types $D$ and $II$ and for type $I$ when none of the
main Weyl scalar invariants vanish. This result is based in a
general expression that relates the $n$-powers of the BR and Weyl
tensors. On the other hand, this expression for $n=2$ is used in the
last section of this work in showing that the minimal polynomial of
the BR tensor distinguishes between types $II$ and $D$.

\subsection{Gravitational superenergy tensors of order n}

The tensor $T^n = {\cal W}^n \diamond \overline{\cal G}$ for $n > 1$
is not, generically, a completely symmetric tensor, because ${\cal
W}^n$ does not inherit the symmetries of ${\cal W}$: it is a
SD-endomorphism but it is not trace-free (in general) and so the
Bianchi identity is not satisfied. But its traceless part
\begin{equation} \label{W(n)}
{\cal W}_{(n)} = {\cal W}^n -  \omega_{(n)} \, {\cal G} \, , \qquad
\omega_{(n)} \equiv \frac13 \Tr {\cal W}^n \, ,
\end{equation}
has the symmetries of the SD Weyl tensor. Thus, it has an associated
superenergy tensor given by:
$$
T_{(n)} \equiv {\cal W}_{(n)} \diamond \overline{{\cal W}}_{(n)} \,
,
$$
that has all the superenergy properties of the BR tensor $T$. We
will say that $T_{(n)}$ is the {\em BR superenergy tensor of order
n}.

All the expression presented in section \ref{sec:BR-e} involving the
Weyl tensor $W$ and the BR tensor $T$ will also be valid for
$W_{(n)}$ and $T_{(n)}$. Specifically,
\begin{equation} \label{T(n)-b}
T_{(n)} =  \Omega_{(n)} \bullet  \overline{{\Omega}}_{(n)}  \, ,
\qquad \Omega_{(n)} \equiv  {\cal W}_{(n)} \diamond \overline{{\cal
G}} \, .
\end{equation}
From lemma \ref{lemma-Omega-n} and (\ref{W(n)}) the last expression
above becomes:
$$
\Omega_{(n)} = \Omega^n -  \omega_{(n)} \, \Gamma \, .
$$
If we use this expression to expand the first expression in
(\ref{T(n)-b}), we obtain the following.
\begin{proposition} The BR superenergy tensor of order n
associated with the Weyl-like tensor ${\cal W}_{(n)}$ defined in
{\rm (\ref{W(n)})} takes the expression:
\begin{equation} \label{T(n)-c}
T_{(n)}  =  T^n - \overline{\omega}_{(n)} \Omega_{(n)} -
\omega_{(n)} \overline{\Omega}_{(n)}  -  |\omega_{(n)}|^2 \ \Gamma
\, .
\end{equation}
\end{proposition}

\subsection{A relation between the $n$-powers of the Weyl and BR tensors}
\label{subsec-Wn}

From the last expression above, we can obtain its self-dual
antisymmetric part by contracting the indexes (23) with the
SD-identity ${\cal G}$. Then, the left side vanishes due to the
whole symmetry of $T_{(n)}$. On the other hand, the right side can
be computed taking into account: first, the property of Appendix
\ref{ap-SD-TLS}.4, second, the definition (\ref{W(n)}), and third,
that $\Omega_{(n)}$ has the same properties as $\Omega$ and,
consequently, it satisfies (\ref{Omega-4}) and (\ref{Omega-6}) by
replacing ${\cal W}$ with ${\cal W}_{(n)}$. Finally, we obtain the
following.
\begin{lemma}
The power n of a SD Weyl tensor is related to the power n of the
associated BR-tensor by
\begin{equation} \label{otra}
2 \, (T^n)_{\alpha \mu \nu \beta} \, {\cal G}^{\mu \nu}_{\ \ \, \rho
\sigma} = \left[ (\Tr \overline{\cal W}^n) \, {\cal W}^n -
\frac{1}{2} | \Tr {\cal W}^n |^2 \, {\cal G}\right]_{\alpha \beta
\rho \sigma}  \, .
\end{equation}
\end{lemma}
When $\Tr {\cal W}^n = 0$, then ${\cal W}^n = {\cal W}_{(n)}$ and
$T^n = T_{(n)}$ is a completely symmetric tensor and, consequently,
(\ref{otra}) is an identity. Nevertheless, when $\Tr {\cal W}^n$
does not vanish, this proposition provides ${\cal W}^n$ from $T^n$,
up to a duality rotation. Indeed, in this case $\Tr T^n = | \Tr
{\cal W}^n |^2 \not= 0$. Then, from (\ref{otra}) we obtain:
\begin{proposition}
If $\, T$ is a BR-tensor and $\Tr T^n \neq 0$, the power n of the
original SD Weyl tensor is given, up to a duality rotation, by
\begin{equation} \label{seisi}
{\cal W}^n = e^{\ci \theta_n} \left[\frac{2}{\sqrt{\Tr{T^n}}} \,
{\cal T}_{(n)} + \frac{\sqrt{\Tr{T^n}}}{2}\, {\cal G}\right] \, ,
\qquad e^{\ci \theta_n} = \frac{\Tr {\cal W}^n}{| \Tr {\cal W}^n |}
\end{equation}
\begin{equation} \label{calT(n)}
{\cal T}_{(n) \alpha \beta \rho
\sigma} \equiv  (T^n)_{\alpha \mu \nu \beta} \, {\cal G}^{\mu
\nu}_{\ \ \, \rho \sigma}
\end{equation}
\end{proposition}

\subsection{The Weyl tensor from the BR tensor}

Let us suppose now $a = \Tr {\cal W}^2 \not=0$ and $b = \Tr {\cal
W}^3 \not=0$. Then, for $n=3$ expression (\ref{seisi}) can be
written as:
\begin{equation} \label{seisia}
\frac{\bar{b}}{|a||b|} \, {\cal W}^3 = \frac{1}{|a||b|}\left[2\,
{\cal T}_{(3)} + \frac{1}{2} |b|^2 \, {\cal G}\right] \, .
\end{equation}
Then, if we remove ${\cal W}^3$ by using the characteristic equation
(\ref{eccar-weyl-2}), and we take into account that $|a|^2 = \Tr
T^2$, $|b|^2 = \Tr T^3$, the SD Weyl tensor can be recovered from
the BR tensor as follows.
\begin{theorem} \label{propdos}
If $T$ is a BR tensor and
   $\Tr T^2 \neq 0$, $\Tr T^3 \neq 0$, then the
  Weyl tensor can be obtained, up to duality rotation, as
\begin{equation} \label{W(T)}
{\cal W} = e^{\ci \theta}{\cal W}_0 \, , \qquad {\cal W}_0 =
\frac{1}{\sqrt{\Tr T^2 \tr T^3}} \left[4 \, {\cal T}_{(3)} +
\frac{1}{3}\, |\Tr T^3|\,{\cal G}\right]
\end{equation}
\begin{equation}
{\cal T}_{(3) \alpha \beta \rho \sigma} \equiv  (T^3)_{\alpha \mu
\nu \beta} \, {\cal G}^{\mu \nu}_{\ \ \, \rho \sigma}
\end{equation}
\end{theorem}

This result partially solves the point (III) stated in the
introduction. It does not apply when $\alpha \beta = 0$, that is,
for the classes $N$, $III$, $IM^{\infty}$ and $IM^{[-6]}$. In these
degenerate cases the determination of the Weyl tensor in terms of
the BR tensor requires a different analysis that will be tackled
elsewhere \cite{fs-BR-can}.

On the other hand, these results provide an alternative approach to
know when a traceless completely symmetric rank 4 tensor $T$ is the
BR tensor associated with a certain Weyl tensor. An answer to this
question has been given by  Bergqvist and Lankien
\cite{bergqvist-lan-1}. From theorem \ref{propdos} an alternative
answer follows when $\alpha $ and $\beta $ do not vanish. Indeed, if
for a given $T$ we calculate the SD double 2-form ${\cal W}_0(T)$ by
using expression (\ref{W(T)}), then $T$ must satisfy the equation $T
= {\cal W}_0(T) \diamond \overline{{\cal W}}_0(T)$.

\section{Complete classification of the Bel-Robinson tensor}
\label{sec-BR-class-b}

At this point we are ready to complete the classification of the BR
tensor by considering the minimal polynomial and not only the
eigenvalue multiplicity (see section \ref{sec-BR-class-a}). Although
a whole analysis of the canonical forms will be tackled elsewhere
\cite{fs-BR-can}, here we will obtain the necessary results to
distinguish between types $N$ and $III$, and between types $II$ and
$D$, and to intrinsically characterize them.

To accomplish this goal, we start from the conditions on the Weyl
tensor that distinguish these Petrov-Bel types (see, for example,
ref. \cite{fms}):
\begin{lemma} \label{lemma-petrov-bel-pm}
Let $a = \Tr {\cal W}^2$, $b = \Tr {\cal W}^3$ the main Weyl scalar
invariants. Then, the Weyl tensor is

i. Petrov-Bel type N iff $\ \ {\cal W}^2 = 0 \neq {\cal W}$.

ii. Petrov-Bel type III iff $\ \ {\cal W}^3 = 0 \neq {\cal W}^2$.

iii. Petrov-Bel type D iff $\ \ a \neq 0 \, , \ {\cal W}^2
=\frac{b}{a} {\cal W} + 2 \frac{b^2}{a^2} {\cal G}$.

\end{lemma}

If we apply the property of Appendix \ref{ap-SD-TLS}.6 to expression
(\ref{Tn-trTn}) we have ${\cal W}^n=0$ if, and only if, $T^n=0$.
Then, considering this property for $n=2,3$, and as consequence of
lemma \ref{lemma-petrov-bel-pm} we can state:
\begin{proposition} \label{prop-I-III}
Let $T$ be a non null BR tensor. Then, the Weyl tensor is:

i. Petrov-Bel type N if, and only if, $\ T^2=0$.

ii. Petrov-Bel type III if, and only if, $\ T^3= 0 \neq T^2$.

\end{proposition}

This proposition shows that Petrov-Bel types $N$ and $III$
correspond to different algebraic classes of the BR tensor because
they have different minimal polynomial. Moreover, it affords a
characterization of these Petrov-Bel types in terms of the sole BR
tensor. The result for type $N$ was given by Bergqvist
\cite{bergqvist} together with the equivalent condition $\,
T_{\alpha \beta \mu \nu} T^{\nu \rho \sigma \lambda} = 0$. This
requisite states, equivalently, that the superenergy flow vector is
a null vector for every observer, a condition that was already
presented as a characterization of the type $N$ Weyl tensors
\cite{bose}. Nevertheless, in this work all the conditions that
characterize the different algebraic classes are written as
equations on the BR tensor as a TLS-endomorphism.

Now let us deal with a type D Weyl tensor. As the SD Weyl tensor
always satisfies the characteristic equation (\ref{eccar-weyl-2}),
we can compute $T^3 = {\cal W}^3 \diamond \overline{\cal W}^3$ to
get:
\begin{equation} \label{auxi1}
T^3 = \alpha^2 T + \beta^2 \Gamma + \frac{1}{6} (a \overline{b}
\Omega + \overline{a} b \overline{\Omega}) \, .
\end{equation}

On the other hand, from the minimal polynomial of a type $D$ Weyl
tensor (see point (iii) of lemma \ref{lemma-petrov-bel-pm}), we can
compute $T^2 = {\cal W}^2 \diamond \overline{\cal W}^2 $ and we
obtain:
\begin{equation} \label{auxi4}
 \alpha T^2 = \frac{\alpha^2}{3} \,  T + \frac{\alpha^3}{3^2} \, \Gamma  +\frac{1}{6}  (
 a \overline{b} \Omega + \overline{a} b \overline{\Omega}), \qquad
 3^3 \beta^2 = 2^2 \alpha^3 \, .
\end{equation}

Then, we can use (\ref{auxi4}) to remove $a \overline{b} \Omega +
\overline{a} b \overline{\Omega} $ in (\ref{auxi1}) and we achieve
that $T$ satisfies a polynomial of degree 3. Moreover, we know from
proposition \ref{prop-BR-eignevalues-b} that, for a type D Weyl
tensor, the BR tensor has 3 different eigenvalues. Thus, this
polynomial is the minimal one. The specific calculation leads to the
following.
\begin{lemma}
 For a type D Weyl tensor, the associated BR tensor $T$ satisfies
 the minimal polynomial:
 \begin{equation} \label{mind}
 T^3 = \alpha T^2 + \frac{2}{3} \alpha^2 T - \frac{2^3}{3^3}
 \alpha^3 \Gamma, \qquad \alpha \neq 0 \, .
 \end{equation}
\end{lemma}

We want to prove now that the converse is also true, that is, that
condition (\ref{mind}) characterize the type D case. Taking the
trace of (\ref{mind}) we obtain $2^2 \alpha^3 = 3^3 \beta^2$.
Moreover, we can use expression (\ref{auxi1}), valid for an
arbitrary BR tensor, to remove $T^3$ in (\ref{mind}), and thus we
obtain that $T$ satisfies (\ref{auxi4}).

On the other hand, expression (\ref{seisi}) for $n=2$ becomes:
$$
{\cal W}^2 = \frac{2a}{|a|^2} {\cal T}_{(2)} + \frac{a}{2} {\cal G}
\, .
$$

Now, if we compute ${\cal T}_{(2)}$ (defined in (\ref{calT(n)}))
putting the expression (\ref{auxi4}) for $T^2$, and we take into
account expressions (\ref{Omega-4}) and (\ref{Omega-6}) and property
of Appendix \ref{ap-SD-TLS}.4, we obtain that the SD Weyl tensor
satisfies an equation of degree 2. Besides, $\alpha$ and,
consequently, $a$ does not vanish. Then, the Weyl tensor is type
$D$.

Moreover, if (\ref{mind}) does not hold and the main BR scalar
invariants satisfy the conditions (i) in proposition
\ref{prop-BR-eignevalues-b}, the Weyl tensor is type $II$. These
scalar conditions can be changed to other equivalent expressions
and, finally, we arrive to the following.
\begin{proposition}  \label{prop-D-II}
Let $T$ be a BR tensor. Then, the Weyl tensor is:

i. Petrov-Bel type D if, and only if, $T$ satisfies {\rm
(\ref{mind})}.

ii. Petrov-Bel type II if, and only if, $\ T$ satisfies  $\ \mu^2= 4
\alpha^6 \beta^4 \not=0$, $2 \mu = 3^3 \beta^4 $ and it does not
satisfy {\rm (\ref{mind})}.

\end{proposition}

Again this proposition shows that Petrov-Bel types $D$ and $II$
correspond to different algebraic classes of the BR tensor because
they have different minimal polynomial.

Propositions \ref{prop-BR-eignevalues-a-0},
\ref{prop-BR-eignevalues-a}, \ref{prop-BR-eignevalues-b},
\ref{prop-BR-eignevalues-c}, \ref{prop-I-III} and \ref{prop-D-II}
afford a characterization of the different algebraic classes of the
BR tensor. These classes correspond to different algebraic classes
of the Weyl tensor, namely, the Petrov-Bel types and some
`degenerate' type $I$ cases. All these result can be summarized by
the following.
\begin{theorem}
For a non vanishing Bel-Robinson tensor $T$ we can distinguish nine
different algebraic classes. The following arrow diagram offers an
algorithm that distinguishes them. In it, $\alpha, \beta, \mu$
denote the main BR scalar invariants defined in {\rm
(\ref{invar-BR})}.
\end{theorem}

\newpage

\vspace*{0.5cm}

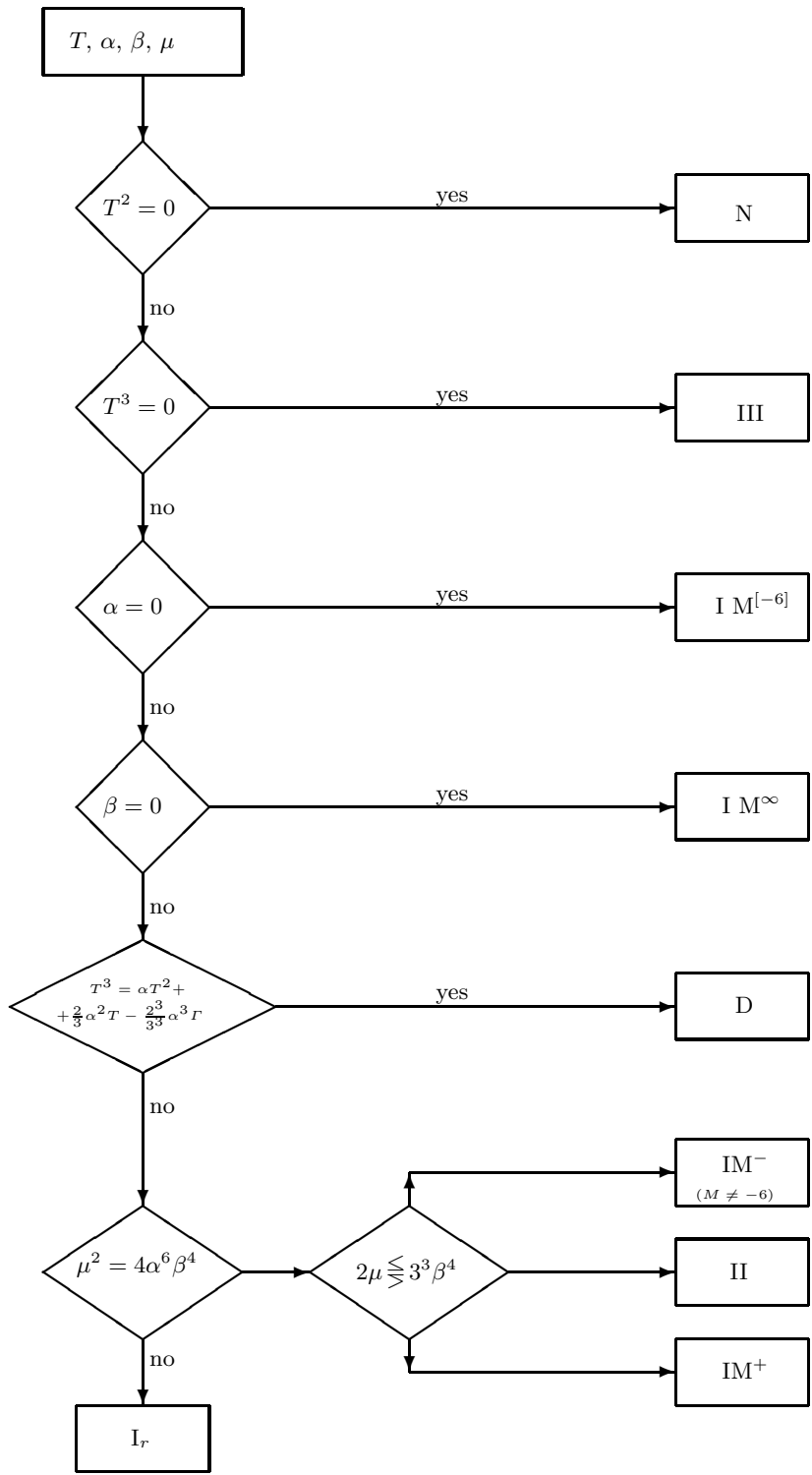
\setlength{\unitlength}{0.9cm} {\footnotesize \noindent
\begin{picture}(0,18)
\thicklines

\put(0.5,18){\line(1,0){3}} \put(0.5,18){\line(0,1){1}}
\put(3.5,19){\line(-1,0){3}} \put(3.5,19){\line(0,-1){1}}
\put(0.9,18.4){$T, \, \alpha  , \,  \beta  , \, \mu  $}

\put(2,18){\vector(0,-1){1}}

\put(2,17){\line(-1,-1){1}} \put(2,17){\line(1,-1){1}}
\put(2,15){\line(1,1){1}} \put(2,15){\line(-1,1){1}}
\put(1.4,15.9){$T^2 =0$}

\put(3,16){\vector(1,0){7}} \put(6.4,16.1){yes}
\put(2,15){\vector(0,-1){1}} \put(2.1,14.4){no}

\put(10.,15.5){\line(0,1){1}} \put(12,15.5){\line(0,1){1}}
\put(10,16.5){\line(1,0){2}} \put(10,15.5){\line(1,0){2}}
\put(10.9,15.8){N }

\put(2,14){\line(-1,-1){1}} \put(2,14){\line(1,-1){1}}
\put(2,12){\line(1,1){1}} \put(2,12){\line(-1,1){1}}
\put(1.4,12.9){$T^3 = 0$}

\put(3,13){\vector(1,0){7}} \put(6.4,13.1){yes}
\put(2,12){\vector(0,-1){1}} \put(2.1,11.4){no}

\put(10,12.5){\line(0,1){1}} \put(12,12.5){\line(0,1){1}}
\put(10,13.5){\line(1,0){2}} \put(10,12.5){\line(1,0){2}}
\put(10.8,12.8){ III }

\put(2,11){\line(-1,-1){1}} \put(2,11){\line(1,-1){1}}
\put(2,9){\line(1,1){1}} \put(2,9){\line(-1,1){1}}
\put(1.4,9.9){$\alpha =0$}

\put(3,10){\vector(1,0){7}} \put(6.4,10.1){yes}

\put(10,9.5){\line(0,1){1}} \put(12,9.5){\line(0,1){1}}
\put(10,10.5){\line(1,0){2}} \put(10,9.5){\line(1,0){2}}
\put(10.6,9.9){I M$^{[-6]}$  }

\put(2,8){\line(-1,-1){1}} \put(2,8){\line(1,-1){1}}
\put(2,6){\line(1,1){1}} \put(2,6){\line(-1,1){1}}
\put(1.4,6.9){$\beta =0$}

\put(3,7){\vector(1,0){7}} \put(6.4,7.1){yes}
\put(2,9){\vector(0,-1){1}} \put(2.1,8.4){no}

\put(10,6.5){\line(0,1){1}} \put(12,6.5){\line(0,1){1}}
\put(10,7.5){\line(1,0){2}} \put(10,6.5){\line(1,0){2}}
\put(10.7,6.9){I  M$^\infty$  }

\put(0,4){\line(2,1){2}} \put(0,4){\line(2,-1){2}}
\put(4,4){\line(-2,1){2}} \put(4,4){\line(-2,-1){2}}
 \put(1.2,4.2){\tiny{$ T^3 =   \alpha   T^2  + $}}

\put(0.7,3.8){\tiny{$+ \frac{2}{3}   \alpha^2 T - \frac{2^3}{3^3}
\alpha^3 \Gamma$}}

\put(10,3.5){\line(0,1){1}} \put(12,3.5){\line(0,1){1}}
\put(10,4.5){\line(1,0){2}} \put(10,3.5){\line(1,0){2}}
\put(10.9,3.9){D }

\put(4,4){\vector(1,0){6}} \put(6.4,4.1){yes}
\put(2,6){\vector(0,-1){1}} \put(2.1,5.4){no}

\put(0.5,0){\line(3,2){1.5}} \put(0.5,0){\line(3,-2){1.5}}
\put(3.5,0){\line(-3,2){1.5}} \put(3.5,0){\line(-3,-2){1.5}}
 \put(1,0){{$ \mu^2 = 4 \alpha^6 \beta^4  $}}

\put(2,3){\vector(0,-1){2}} \put(2.1,2.4){no}

 \put(6,1){\line(-3,-2){1.5}} \put(6,1){\line(3,-2){1.5}}
\put(6,-1){\line(3,2){1.5}} \put(6,-1){\line(-3,2){1.5}}
\put(5.2,-0.1){$ 2\mu    \begin{array}{l} < \\[-2mm] = \\[-2mm] > \end{array}
3^3 \beta^4 $}

\put(10,1){\line(0,1){1}} \put(12,1){\line(0,1){1}}
\put(10,2){\line(1,0){2}} \put(10,1){\line(1,0){2}}
\put(10.7,1.5){IM$^- $  } \put(10.3,1.1){\tiny{($M\neq-6$)} }

\put(10,-2){\line(0,1){1}} \put(12,-2){\line(0,1){1}}
\put(10,-1){\line(1,0){2}} \put(10,-2){\line(1,0){2}}
\put(10.7,-1.6){IM$^+ $ }

\put(10,-0.5){\line(0,1){1}} \put(12,-0.5){\line(0,1){1}}
\put(10,0.5){\line(1,0){2}} \put(10,-0.5){\line(1,0){2}}
\put(10.7,-0.1){ II }

\put(7.5,0){\vector(1,0){2.5}}

\put(6,1.5){\vector(1,0){4}}

\put(6,-1.5){\vector(1,0){4}}

\put(6,-1){\vector(0,-1){0.5}} \put(6,1){\vector(0,1){0.5}}

\put(3.5,0){\vector(1,0){1}} \put(2,-1){\vector(0,-1){1}}

\put(1,-3){\line(0,1){1}} \put(3,-3){\line(0,1){1}}
\put(1,-2){\line(1,0){2}} \put(1,-3){\line(1,0){2}} \put(1.7,-2.6){
I$_r$  }

 \put(2.1,-1.4){no}

\end{picture}

}

\newpage

\begin{acknowledgements}
The authors would like thank J. A. Morales for some useful comments.
This work has been partially supported by the Spanish Ministerio de
Educaci\'on y Ciencia, MEC-FEDER project FIS2006-06062.
\end{acknowledgements}

\appendix

\section{Products and other formulas involving 2-tensors
$A$ and $B$} \label{ap-2tensors}
\begin{enumerate}
\item Composition as endomorphisms: $A \cdot B $,
$$ {(A \cdot B)^{\alpha}}_{\beta}= {A^{\alpha}}_{\mu}
{B^{\mu}}_{\beta}$$
\item In general, for arbitrary tensors, $\cdot$ will be used to indicate the
contraction of adjacent indexes on the tensorial product.
\item Square and trace as endomorphism
$$A^2 = A \cdot A, \qquad \tr A = {A^{\alpha}}_{\alpha}$$
\item Action on vectors $x,\, y$ as an endomorphism $A(x)$ and as a bilinear form $A(x,y)$:
$$A(x)^{\alpha} ={A^{\alpha}}_{\beta} x^{\beta}, \qquad A(x,y) =
A_{\alpha \beta} x^{\alpha} y^{\beta} $$
\item Exterior product as double 1-forms: $A \wedge B$,
$$( A \wedge B)_{\alpha \beta \mu \nu} =
A_{\alpha \mu} B_{\beta \nu} + B_{\alpha \mu} A_{\beta \nu} -
A_{\alpha \nu} B_{\beta \mu} -B_{\alpha \nu} A_{\beta \mu}$$

\end{enumerate}

\section{Products and other formulas involving SD-endomorphisms
${\cal X}$ and ${\cal Y}$} \label{ap-SD}
\begin{enumerate}
\item Every self-dual (SD) symmetric double 2-form ${\cal X}$ defines a linear map on
the 3-dimensional SD bivector space. For short, we will say that
${\cal X}$ is a SD-endomorphism.
\item Composition as endomorphisms ${\cal X} \circ {\cal Y}$:
$$({\cal X} \circ {\cal Y})_{\alpha \beta \rho \sigma} = \frac{1}{2} \,
{{\cal X}^{\alpha \beta}}_{\mu \nu} {{\cal Y}^{\mu \nu}}_{\rho
\sigma}$$
\item{Square and trace as endomorphism}:
$$ {\cal X}^2= {\cal X} \circ {\cal X}, \qquad \Tr{\cal X} = \frac{1}{2}
{\cal X}^{\alpha \beta}_{\ \ \ \alpha \beta} $$
\item Action (on SD bivectors ${\cal F}$, ${\cal
H}$) as an endomorphism ${\cal X}({\cal F})$, and as a bilinear form
${\cal X}({\cal F},{\cal H})$:
$${\cal X}({\cal F})_{\alpha \beta} = \frac{1}{2} {{\cal X}_{\alpha \beta}}^{\mu \nu} \,
{\cal F}_{\mu \nu}, \qquad {\cal X}({\cal F},{\cal H})= \frac{1}{4}
{\cal X}_{\alpha \beta \mu \nu} {\cal F}^{\alpha \beta} {\cal
H}^{\mu \nu}$$
\item Metric on the space of SD bivectors (SD-identity):
$${\cal G}= \frac{1}{2} ( G - \ci \eta) \, , \qquad {\cal G}({\cal F}) = {\cal F}$$
\item
The SD-identity ${\cal G}$ satisfies:
$${\cal G}_{\alpha \mu \beta \nu} {{\cal G}^{\mu \nu}}_{\rho \sigma} = -
\frac{1}{2}\, {\cal G}_{\alpha \beta \rho \sigma} , \qquad
\overline{\cal G}_{\alpha \mu \beta \nu} {{\cal G}^{\, \mu
\nu}}_{\rho \sigma}= \frac{3}{2}\, {\cal G}_{\alpha \beta \rho
\sigma}$$
\item The (1,3)-trace of a SD-endomorphism ${\cal X}$ is proportional to the
metric. More precisely:
$$
{\cal X}^{ \mu}_{\ \alpha \mu \beta } = \frac12 \Tr {\cal X} \,
g_{\alpha \beta}
$$
\item A SD-endomorphism ${\cal X}$ is traceless iff it satisfies the algebraic
Bianchi identity. More precisely:
$$
\Tr {\cal X} = 0 \quad \Leftrightarrow \quad {\cal X}^{ \mu}_{\
\alpha \mu \beta } = 0  \quad \Leftrightarrow \quad {\cal X}_{\alpha
(\beta \mu \nu)} = 0 \,
$$
\item The metric volume element $\eta$ is a linear map on the
2-forms space that defines the Hodge dual operator. For a real
2-from $F$ and a real symmetric double 2-form $W$:
$$*F = \eta(F) \, , \qquad *W = \eta \circ W \, .$$
\end{enumerate}

\section{Products and other formulas involving TLS-endomorphisms $T$ and $R$.}
\label{ap-TLS}
\begin{enumerate}
\item Every 4-tensor $T$ with the symmetries:
\begin{equation} \label{TLS-def}
T_{\alpha \beta \mu \nu} = T_{\beta \alpha \mu \nu} = T_{\mu \nu
\alpha \beta} \, , \qquad  T^{\alpha}_{\ \alpha \mu \nu} = 0
\end{equation}
defines a symmetric linear map on the 9-dimensional space of the
traceless symmetric tensors. We say that $T$ is a TLS-endomorphism.
\item Composition as endomorphisms: $T  \bullet  R$,
$${(T \bullet R )^{\alpha \beta}}_{\rho \sigma} = {T^{\alpha \beta}}_{\mu \nu}
{R^{\mu \nu}}_{\rho \sigma}$$
\item{Square and trace as endomorphism}:
$$T^2 = T \bullet T \, , \qquad  \Tr T = T^{\alpha \beta}_{\ \ \ \alpha \beta} $$
\item Action (on trace-less symmetric tensors $A$, $B$) as an endomorphism $T(A)$
and as a bilinear form  $T(A, B)$,
$$T(A)_{\alpha \beta} = {T_{\alpha \beta}}^{\mu \nu} \, A_{\mu \nu}, \qquad T(A,B) =
T_{\alpha \beta \mu \nu} \, A^{\alpha \beta} B^{\mu \nu} $$
\item Metric on the space of traceless symmetric tensors (TLS-identity): $\Gamma$
$$\Gamma_{\alpha \beta \mu \nu} =\frac{1}{2} ( g_{\alpha \mu}  g_{\beta \nu} +
g_{\alpha \nu} g_{\beta \mu } ) - \frac{1}{4} g_{\alpha \beta}
g_{\mu \nu} \, , \qquad  \Gamma(A) = A$$
\end{enumerate}

\section{Properties involving TLS-endomorphisms defined by SD-endomorphisms}
\label{ap-SD-TLS}
\begin{enumerate}
\item For two SD-endomorphisms ${\cal X}$ and ${\cal Y}$, we shall define
the $\diamond-$product, ${\cal X} \diamond \overline{{\cal Y}}$ as:
$$({\cal X} \diamond \overline{{\cal Y}})_{\alpha \mu \beta \nu} =
{{{\cal X}_{\alpha}^{\ \ \rho}}_{  \beta}}^{\sigma} \,
\overline{{\cal Y}}_{\mu \rho \nu \sigma}$$
\item If ${\cal X}$ and ${\cal Y}$ are SD-endomorphisms, then
$Q = {\cal X} \diamond \overline{{\cal Y}}$ is a TSL-endomorphism
such that:
$$
Q^{ \mu}_{\ \alpha \mu \beta } = \frac14 \Tr Q \, g_{\alpha \beta}
\, , \qquad \Tr Q = \Tr {\cal X} \, \overline{\Tr {\cal Y}}
$$
\item The TLS-identity $\Gamma$ is given in terms of the SD-identity ${\cal G}$ by:
$$\Gamma = {\cal G} \diamond \overline{\cal G} $$
\item The TLS-identity $\Gamma$ satisfies:
$$
\Gamma_{\alpha \mu \nu \beta}\, {{\cal G}^{\mu \nu}}_{\rho \sigma} =
-\frac34 \, {\cal G}_{\alpha \beta \rho \sigma}
$$
\item If $Q = {\cal X} \diamond \overline{{\cal G}}$, ${\cal
X}$ being a traceless SD-endomorphism, then:
$$
Q_{\alpha \mu \nu \beta}\, {{\cal G}^{\mu \nu}}_{\rho \sigma} = 2 \,
Q_{\alpha [\rho \sigma] \beta} = \frac32 \, {\cal X}_{\alpha \beta
\rho \sigma}
$$
The last two properties follow from identities of Appendix
\ref{ap-SD}.6.
\item If ${\cal X}$ is a SD-endomorphism, then
$$
Q = {\cal X} \diamond \overline{{\cal X}} = 0 \quad \Leftrightarrow
\quad {\cal X} = 0
$$
\item The SD-endomorphism ${\cal X}$ is traceless iff
$Q = {\cal X} \diamond \overline{{\cal X}}$ is a real fully
symmetric traceless tensor.\\
This property can be easy shown by using the properties 2 above and
8 of Appendix \ref{ap-SD}.
\item If ${\cal V}$, ${\cal Z}$, ${\cal X}$ and ${\cal Y}$ are SD-endomorphisms, then:
\begin{equation} \label{ap-D4}
({\cal V} \diamond \overline{{\cal Z}}) \bullet ({\cal X} \diamond
\overline{{\cal Y}}) = ({\cal V} \circ {\cal X}) \diamond
(\overline{{\cal Z} \circ  {\cal Y}})
\end{equation}
This property can be shown by writing in a normalized basis $\{
{\cal U}_i \}$ of the SD bivectors the four SD-endomorphisms
involved,
$$
{\cal V}=\sum_{ij} V_{ij} \, {\cal U}_i \otimes {\cal U}_j \, ,
$$
and computing the two parts of the expression taking into account
that
$$
{\cal G}({\cal U}_i, {\cal U}_j ) = - \delta_{ij} , \quad \sqrt{2} \
{\cal U}_i \cdot {\cal U}_j = - \ci \  \epsilon_{ijk} {\cal U}_k,
\quad {\cal G}({\cal U}_i, \overline{{\cal U}}_j ) = 0 \, .
$$

\end{enumerate}


\begin{thebibliography}{99}



\bibitem{bel-1} Bel, L.: C. R. Acad. Sci. {\bf 247}, 1094 (1958)

\bibitem{bel-2} Bel, L.: C. R. Acad. Sci. {\bf 248}, 1297 (1959)

\bibitem{bel-3} Bel, L.: Cah. de Phys. {\bf 16}, 59 (1962). This article
has been reprinted in Gen. Rel. Grav. {\bf 32}, 2047 (2000)

\bibitem{seno} Senovilla, J.M.M.: Class. Quantum Grav. {\bf 17}, 2799 (2000)

\bibitem{garcia} Garc\'{\i}a-Parrado, A.: Class. Quantum Grav. {\bf 25}, 015006 (2008)

\bibitem{debever-1} Debever, R.: Bulletin de la Societ\'e Math\'ematique de Belgique
{\bf t. X}, 112 (1958)

\bibitem{debever-2} Debever, R.: C. R. Acad. Sci. {\bf 249}, 1324 (1959)

\bibitem{penri-1} R. Penrose and W Rindler, {\it Spinors and
spacetime} (Cambridge U.P., Cambridge, 1984) Vol. 1

\bibitem{bergqvist-lan-1} Bergqvist, G., Lankinen, P.: Class. Quantum Grav. {\bf 21},
3499 (2004)

\bibitem{rai} Rainich, G.Y.: Trans. Am. Math. Soc. {\bf 27}, 106
(1925)

\bibitem{fsY} Ferrando, J.J., S\'aez, J.A.: Gen. Relativ. Gravit. {\bf 35},
1191 (2003)

\bibitem{petrov-W} Petrov, A.Z.: Sci. Not. Kazan Univ. {\bf 114}, 55 (1954). This article
has been reprinted in Gen. Rel. Grav. {\bf 32}, 1665 (2000)

\bibitem{fs-BR-can} Ferrando, J.J., S\'aez, J.A.: (in preparation)

\bibitem{mcar} McIntosh, C.B.G., Arianrhod, R.: Class. Quantum Grav.
{\bf 7}, L213 (1990)

\bibitem{fs-aligned-em} Ferrando, J.J., S\'aez, J.A.: Gen. Relativ. Gravit. {\bf 36},
2497 (2004)

\bibitem{bose} Bonilla, M.À.G., Senovilla, J.M.M.: Phys. Rev. Lett.
{\bf 11}, 783 (1997)

\bibitem{bergqvist} Bergqvist, G.: J. Math. Phys. {\bf 39}, 2141
(1998)

\bibitem{kra} Stephani, H., Kramer, D., MacCallum, M., Hoenselaers, C., Herlt, E.:
{\it Exact solutions to Einstein's field equations} (Cambridge
University Press, Cambridge, 2003)

\bibitem{fms} Ferrando, J.J., Morales, J.A., S\'aez,
J.A.: Class. Quantum Grav. {\bf 18}, 4969 (2001)

\bibitem{fsI} Ferrando, J.J., S\'aez, J.A.: Class. Quantum Grav. {\bf 14}, 129 (1997)

\bibitem{fsWem} Ferrando, J.J., S\'aez, J.A.: Class. Quantum Grav. {\bf 19}, 2437 (2002)



\end{thebibliography}
\end{document}